\documentclass[reprint,pra,amsmath,amssymb,aps,superscriptaddress,floatfix]{revtex4-2}
\usepackage[utf8]{inputenc}
\usepackage[T1]{fontenc}
\usepackage{graphicx}
\graphicspath{{Figures/}{./}}
\usepackage{placeins} 
\usepackage{xcolor}

\newcommand{\shunsuke}[1]{\textcolor{black}{#1}}

\usepackage{hyperref}

\usepackage{float}
\usepackage{textgreek}
\newcommand{\Ec}{\ensuremath{\mathrm{E_c}}}          
\newcommand{\Eeff}{E^{*}}              

\newcommand{\ladhyx}{Laboratoire d’Hydrodynamique (LadHyX), CNRS, École Polytechnique, Institut Polytechnique de Paris, 91120 Palaiseau, France}
\newcommand{\pasteur}{Institut Pasteur, Université Paris Cité, Physical microfluidics and Bioengineering, 25-28 Rue du Dr Roux, 75015 Paris, France}
\newcommand{\kyotomech}{Mechanical and Systems Engineering Undergraduate Course Program, Undergraduate School of Engineering Science, Kyoto University, Kyoto 615-8540, Japan.}

\begin{document}
\title{Deformation and organization of droplet-encapsulated soft beads}

\author{Shunsuke Saita}
\affiliation{\ladhyx}
\affiliation{\kyotomech}

\author{Finn Bastian Molzahn}
\affiliation{\ladhyx}

\author{Clara Delahousse}
\affiliation{\ladhyx}
\affiliation{\pasteur}

\author{Julien Husson}
\affiliation{\ladhyx}

\author{Charles N. Baroud}
\email{charles.baroud@polytechnique.edu}
\affiliation{\ladhyx}
\affiliation{\pasteur}

\date{\today}

\begin{abstract}
Many biological, culinary, and engineering processes lead to the co-encapsulation of several soft particles within a liquid interface. In these situations the particles are bound together by the capillary forces that deform them and influence their biological or rheological properties. Here, we introduce an experimental approach to encapsulate a controlled number of soft beads within aqueous droplets in oil. These droplet-encapsulated gels are manipulated in a deformable microfluidic device to merge them and modify the liquid fraction. In the dry limit the contact surface between the hydrogels is found to be determined by the elastocapillary number $E_c$, with the contact radius following a $E_c^{1/3}$ dependence,  indicating that the deformation increases for soft or small particles. When multiple beads are co-encapsulated within a single droplet they can be arranged into linear or three-dimensional aggregates that remain at a local energy minimum.
\end{abstract}

\maketitle


The interaction of soft particles within liquid droplets is a recurring motif in both engineering applications and natural phenomena. This is the case in molecular cuisine based on soft gel beads ~\cite{yuasa2019texture, brandao2023molecular} or in industrial spray-drying processes, where edible particles (e.g.  powdered milk or fruit purees \cite{caric1987effects, caparino_effect_2012}) are co-encapsulated within shrinking liquid volumes~\cite{muhoza2023spray}. In parallel, ecological and biological situations can involve gel-like particles being joined together with an air-liquid interface, such as the case of frog eggs on the surface of a pond (frog spawn~\cite{frogspawn}). \shunsuke{Similar physics can be encountered in recent technological applications, including the co-encapsulation of multiple microparticles within microfluidic droplets for single-cell genomic analysis~\cite{delley2020microfluidic, du_deterministic_2025}, or 3D printing of soft particles~\cite{Roh2017Capillary}.} 

In all of the above cases, a force due to the surface tension binds the soft particles together and deforms them, with implications on the mechanical stability of the aggregates, the rheology of the foods, or the biological outcome for the cells. \shunsuke{But while the deformation of soft particles due to external forces or adhesion between them is a well-studied problem (see e.g. Refs.~\cite{Ramachandran2011Adhesive,Dintwa2008Accuracy,Ciavarella2019Role}), linking these deformations with surface tension forcing (elastocapillarity) must account for shape and volume changes in both the solid and liquid media. This elastocapillary balance is well understood for 1D fibers or 2D sheets~\cite{style_elastocapillarity_2017,bico_elastocapillarity_2018} but it has received much less attention in 3D, where most of the studies focused on the wetting on flat surfaces in the presence of elastic effects~\cite{jensen_wetting_2015,chakrabarti_elastowetting_2018,chen_soft_2024}. }

Here, we study the energy balance that determines the shape of two soft beads under the action of interfacial tension by introducing droplet-encapsulated gels (DEGs): hydrogel bead pairs or clusters co-residing within a single droplet. We begin by determining the dominant physical ingredients that dictate the shape of the two-bead system, before exploring more complex situations with multiple beads.

\begin{figure}[ht]
  \centering
  \includegraphics[width=0.7\columnwidth]{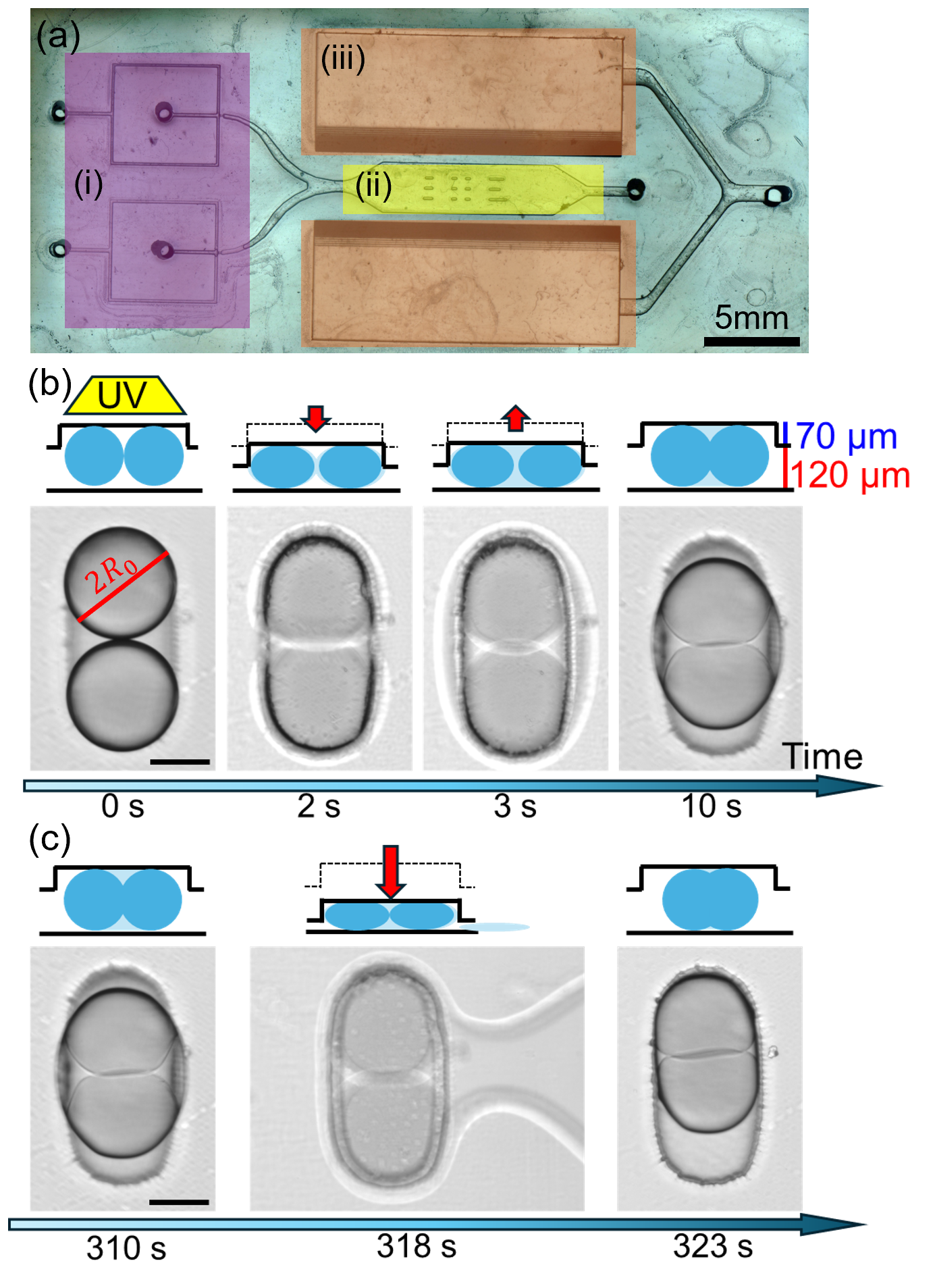}
  \caption{  (a) Droplets are produced at flow-focusing junctions (i), trapped in capillary anchors within the observation chamber (ii), and the chamber ceiling is lowered by evacuating air from side cavities (iii). 
  (b) Ultraviolet exposure polymerizes the droplets into gel beads; subsequent ceiling actuation compresses the bead pairs and the expelled liquid coalesces into an outer droplet. 
  (c) Additional compression decreases the encapsulating liquid volume. 
  All scale bars, $100~\mu\mathrm{m}$.}
\label{fig:method}
\end{figure}

The experiments were conducted in custom polydimethylsiloxane (PDMS) microfluidic channels comprising three main elements (Fig.~\ref{fig:method}a). Two flow-focusing junctions \cite{anna_formation_2003} produced two independent droplet types (Fig.~\ref{fig:method}a(i)). The droplets then entered an observation chamber and were trapped in capillary anchors \cite{dangla_trapping_2011} patterned into the chamber ceiling (Fig.~\ref{fig:method}a(ii)). A separate fluidic circuit connected two large air chambers to an external pressure source (Fig.~\ref{fig:method}a(iii)), such that applying negative (positive) pressure to these chambers lowered (raised) the ceiling height in the observation chamber \cite{jain_using_2024}. The device was fabricated by molding PDMS on a 3D-printed master (ASIGA MAX X, PlasGray V2 resin), rendering surfaces hydrophobic with NOVEC~1700, and casting/bonding PDMS to complete the chip. PEGDA (poly(ethylene glycol) diacrylate) droplets were generated at the flow-focusing junctions, with FC-40 containing 0.5\% (v/v) fluorosurfactant (FluoSurf) as the continuous phase. 

The experimental protocol began by generating liquid droplets and capturing them in the capillary anchors, after which the flow rate of the continuous phase was increased to flush away untrapped droplets. This procedure yielded linear pairs or clusters of $N=3,4,5,6$ droplets within each anchor, depending on the anchor length. Each droplet was then individually cross-linked through exposure to ultraviolet light for $5\,\mathrm{s}$, using the DAPI fluorescence filter and a 40x objective.  \shunsuke{Since the photo-polymerization of PEG-DA is inhibited by the presence of oxygen, a thin layer of unpolymerized gel remains around the gel bead after illumination~\cite{Xia2016Oxygen}. Nevertheless, the crosslinking is expected to yield homogeneous mechanical properties in the bulk of the beads~\cite{dendukuri_modeling_2008} (See SI for detailed calculation~\cite{saita_supplementary}).}

{Following this cross-linking step, the ceiling of the observation chamber was lowered by aspirating the air from the side cavity using a $25\,\mathrm{ml}$ glass syringe at a rate of around $2-3$~{ml/s} (See Fig.~\ref{fig:method}b). During this compression the droplets remained trapped in the microfluidic anchors due to a combination of capillary and elastic resistance~\cite{tomasi_individual_2020}. The beads however expelled part of their internal liquid into a surrounding puddle and the expelled liquid from adjacent beads merged together, forming a doublet of beads within a single aqueous droplet. The aspiration was then reversed at approximately the same rate, leaving the two beads (diameters in the range $150-172~\mu$m) in contact due to surface tension but unconfined by the channel ceiling.} 

{Further repeating the ceiling compression after the initial merging was then used to expel some of the liquid contained in the DEG (see Fig.~\ref{fig:method}c), which allowed us to reduce the liquid fraction $\phi$ within the DEGs. After each deformation step, a minimum wait time of $5\,\mathrm{min}$ was allowed for the liquid to equilibrate within the hydrogel network~\cite{louf2021poroelastic}. Measurements of the bead sizes showed that the bead volumes remained nearly constant until $\phi\simeq 0.15$, then gradually dropped as $\phi$ was further decreased. The maximum volume reduction of the gel beads was in the range of 9-22\% for $\phi\rightarrow0$ (SI Fig.~S1). }

\begin{figure}[ht]
  \centering
  \includegraphics[width=\columnwidth]{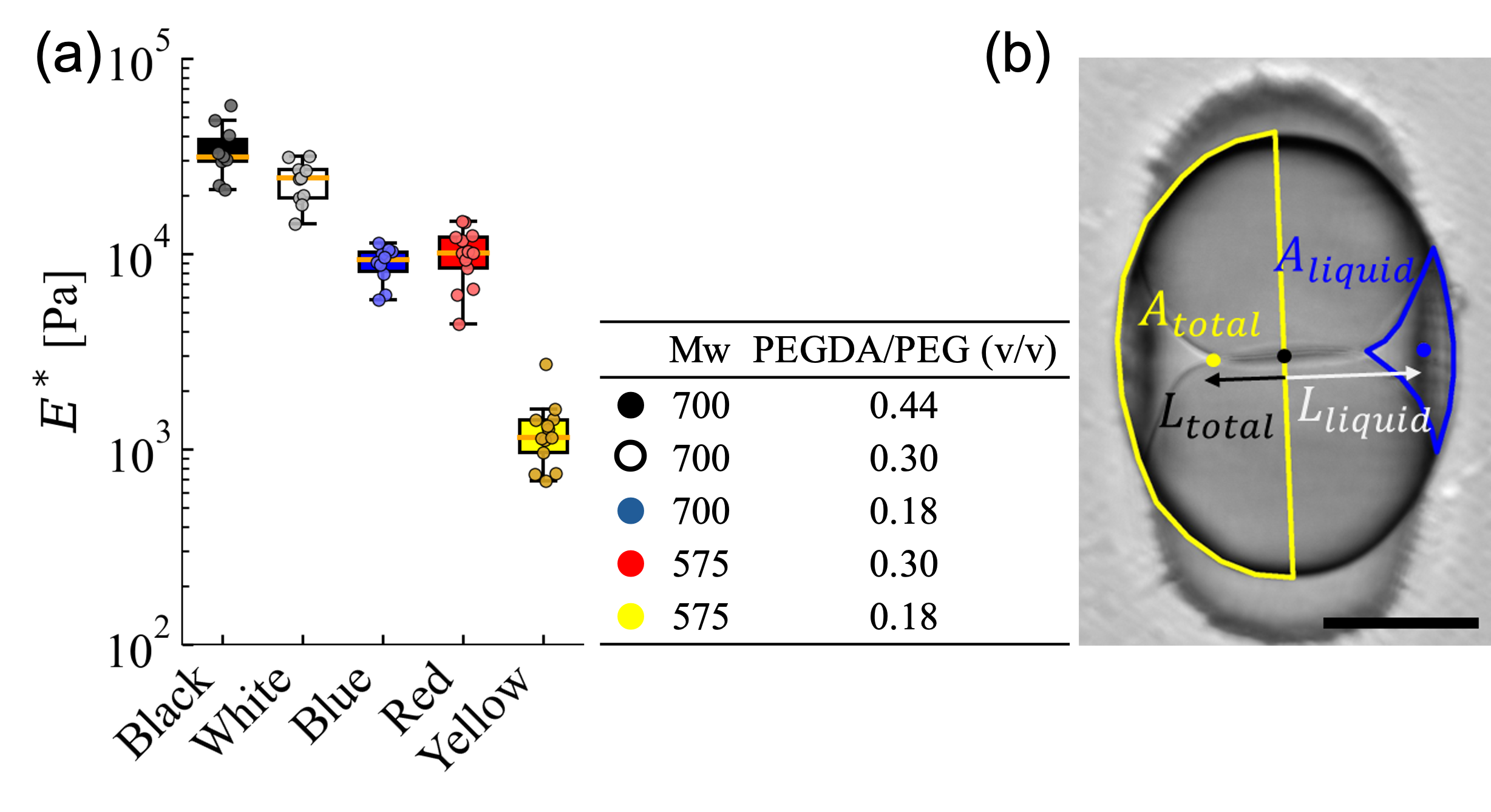}
  \caption{(a) Effective elasticity $E^{*}$ of five PEGDA formulations, measured by micropipette indentation. The table presents the PEGDA molecular weight and the PEGDA:PEG volume ratio for each color code. 
 (b) Image-based measurement of the liquid volume fraction $\phi$ in Fiji~\cite{schindelin_fiji_2012}. Cross-sectional areas ($A_{\mathrm{liquid}}$, $A_{\mathrm{total}}$) and centroidal distances to the symmetry axis ($L_{\mathrm{liquid}}$, $L_{\mathrm{total}}$) are obtained from traced polygons under an axial-symmetry assumption, and volumes are computed via Pappus’ theorem. Scale bar $100~\mu\mathrm{m}$.}
  \label{fig:props}
\end{figure}

The shape of the DEG is determined by an equilibrium between the water-oil interfacial tension, which pushes the beads together, and the elastic response of the beads, which resists deformation. The balance between these two physical effects is quantified by calculating the dimensionless elastocapillary number~\cite{bico_elastocapillarity_2018},
\begin{equation}
E_c = \frac{\gamma}{E^{*} D},
\label{eq:elasto-capillary}
\end{equation}
where $\gamma$ is the water–oil interfacial tension, $E^{*}$ is the effective elastic modulus of the bead, and $D$ is the bead diameter. To modify the value of $E_c$, we first tuned the beads’ effective elasticity $E^{*}$ by preparing formulations of PEGDA having different molecular weight and by varying the PEGDA:PEG volume ratio. Five different bead stiffnesses were obtained, which we refer to with a color code (black = stiffest; yellow = softest), as shown in Fig.~\ref{fig:props}a.

The value of $E^{*}$ was measured for each bead condition by two independent methods: a microfluidic deformation assay, using the geometry developed in Ref.~\cite{moore_clogging_2023} (see {SI Fig. S2}~\cite{saita_supplementary}). These measurements were confirmed using a micro-pipette based indentation method \cite{zaidel-bar_measuring_2023} (Fig.~\ref{fig:props}a). The two methods provided similar values, except for the black beads for which the microfluidic assay under-resolved small strains. In the following we therefore use the micro-indentation values for $E^{*}$. The oil–water interfacial tension was measured by pendant-drop tensiometry \cite{berry2015measurement} as $\gamma = 7.9 \pm 1.5~\mathrm{mN\,m^{-1}}$ under our surfactant conditions. 

In addition to $E_c$, the contact strength between two DEGs is also determined by the liquid volume fraction $\phi \equiv V_{\mathrm{liquid}}/V_{\mathrm{total}}$, where $V_{\mathrm{liquid}}$ is the volume of the surrounding liquid (excluding the gels) and $V_{\mathrm{total}}$ is the combined volume of liquid and gels (see Fig.~\ref{fig:props}b). The value of $\phi$ was estimated from the 2D microscopy images by assuming axial symmetry and using Pappus’ centroid theorem. To this end we manually traced the polygonal outline of one half of the liquid region to obtain its cross-sectional area $A_{\mathrm{liquid}}$ and its centroidal distance to the symmetry axis $L_{\mathrm{liquid}}$, yielding $V_{\mathrm{liquid}} = 2\pi A_{\mathrm{liquid}} L_{\mathrm{liquid}}$. Applying the same procedure to the entire cross-section gives $V_{\mathrm{total}} = 2\pi A_{\mathrm{total}} L_{\mathrm{total}}$. Both the left and right halves were traced independently and the two estimates were averaged.

\begin{figure}[ht]
  \centering  \includegraphics[width=0.85\columnwidth]{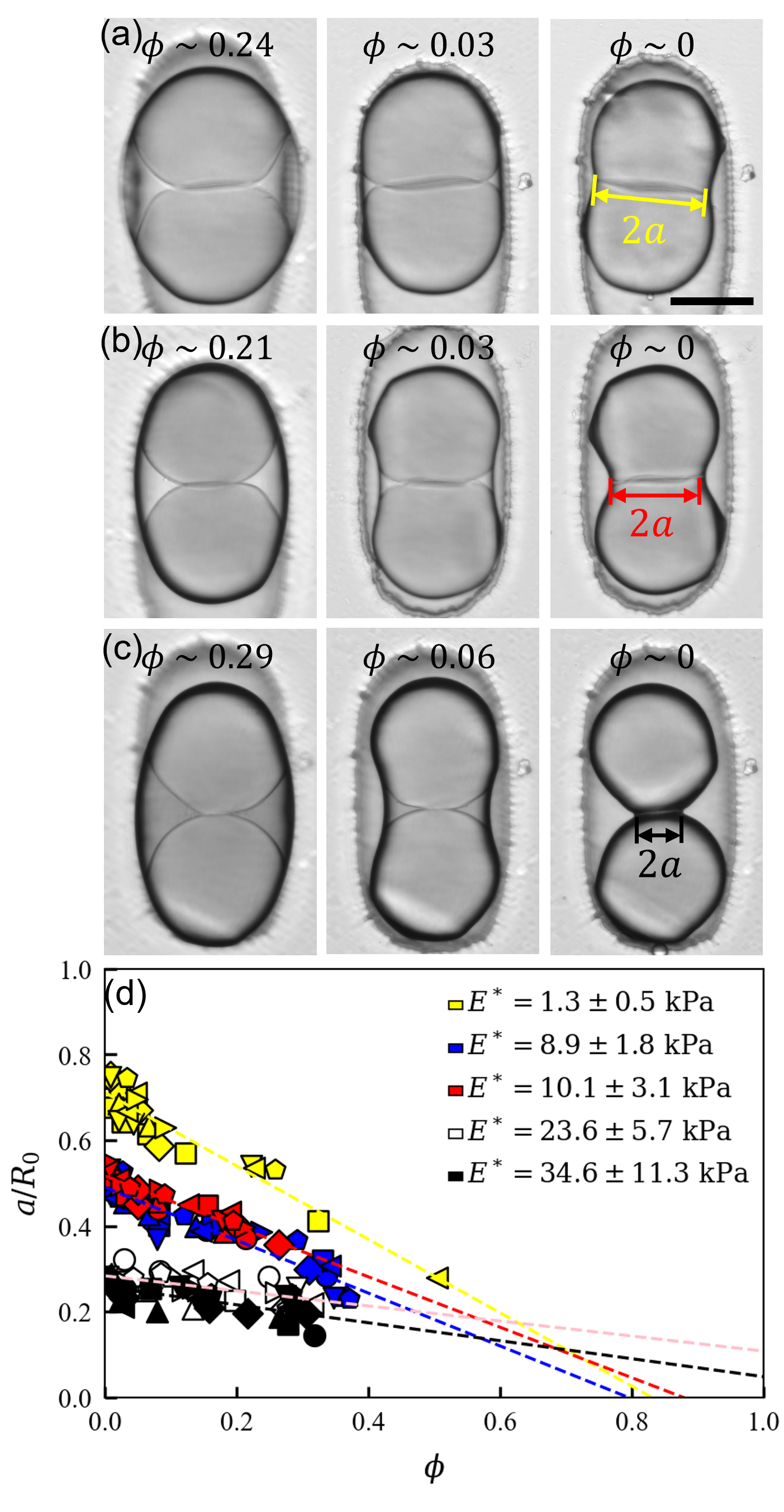}
  \caption{DEGs morphologies shown for different values of  $\phi$ and $\Ec$. 
  (a) Soft gel beads ($\Eeff = 1.3\pm0.5$~kPa);  
  (b) Intermediate gel beads ($\Eeff=10.1\pm3.1$~kPa);  
  (c) Stiff gel beads ($\Eeff=34.6\pm11.3$~kPa). In each row, decreasing $\phi$ from left to right expands the contact length. Scale bar, $100~\mu\mathrm{m}$. 
  (d) Normalized contact radius $a/R_0$ vs. liquid volume fraction $\phi$ for five values of $\Eeff$. The analysis was performed on 8 DEGs. Lines are least-squares linear fits for each gel type within the explored range. \shunsuke{Value of $\Eeff$ given as mean $\pm$ standard deviation.}}
  \label{fig:N2}
\end{figure}
Images for three bead pairs with different compositions are shown in Fig.~\ref{fig:N2}, as $\phi$ is reduced towards $\phi\simeq0$. While the shapes are qualitatively similar for large values of $\phi$, they display strong differences for small volume fractions. These differences can be quantified by measuring the radius $a$ of the contact circle between the two beads and normalizing \shunsuke{by the initial radius of the droplet $R_0$}. The evolution of $a/R_0$ follows a linear trend in $\phi$ for each value of $\Eeff$, as shown by the good agreement between the plotted data and the least-squares linear fits (Fig.~\ref{fig:N2}d).

\begin{figure}[ht]
  \centering  \includegraphics[width=\columnwidth]{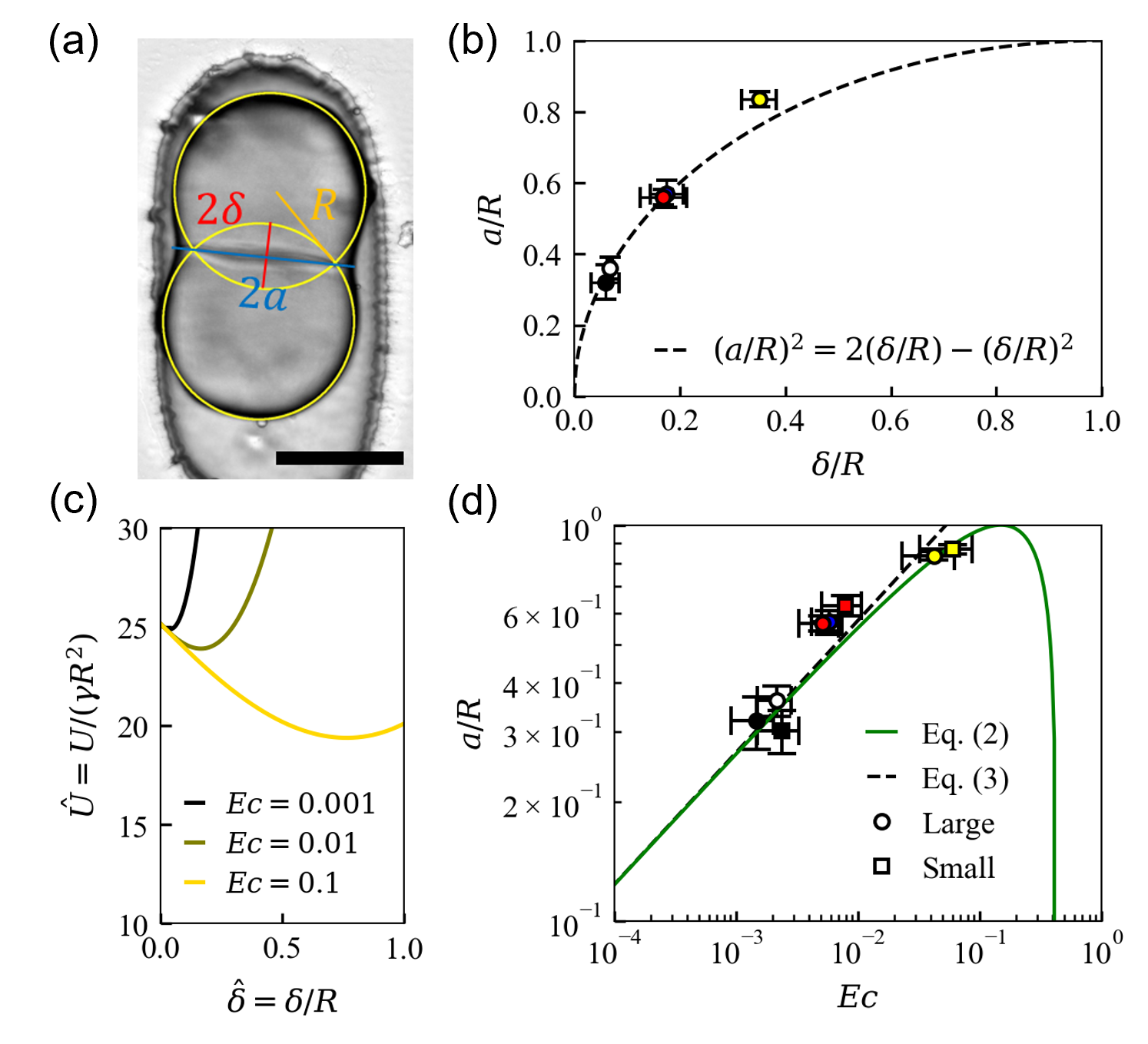}
  \caption{(a) Definition of the geometric parameters $R$, $\delta$, and $a$ and their calculation from experimental images of the DEGs in the dry limit. Scale bar, $100~\mu\mathrm{m}$.
  (b) Comparison of measured $a/R$ versus $\delta/R$ with the geometric relation $a^2 = 2R\delta - \delta^2$ \shunsuke{(data: mean $\pm$ standard deviation)}. 
  (c) Dimensionless energy landscape $\hat{U}(\hat{\delta})$ with $\hat{\delta}=\delta/R$ exhibiting a single minimum for each value of $E_c$ 
  (d) Dimensionless contact length $a/R$ versus elastocapillary number $E_c$ in the dry limit. \shunsuke{Data points correspond to the mean of $n=8$, $4$, $8$, $9$, and $8$ beads for the black, white, blue, red, and yellow large beads, respectively ($D=153~\mu\mathrm{m}$), and $n=4$, $5$, and $5$ beads for the black, red, and yellow small beads, respectively ($D=100~\mu\mathrm{m}$). The vertical error bars represent the standard deviation on the measured data. Horizontal error bars calculated from error propagation.}}
  \label{fig:scaling}
\end{figure}

The evolution of $a$ for different values of $E_c$ can be understood by focusing on the dry limit ($\phi\rightarrow0$). For $\phi = 0$ the DEG geometry reduces to two connected spherical caps, whose interfacial area is $S=2\times(4\pi R^{2}-2\pi R\delta)$, where $R$ is now the measured radius of each of the beads for a value of $\phi$ and $\delta$ represents the indentation (see Fig.~\ref{fig:scaling}a). In the limit of small deformations the Pythagorean theorem provides the geometric relation $a^{2} = 2R\delta - \delta^{2}$, while very soft beads are expected to depart from this relation. Testing the relation between $a$ and $\delta$ therefore serves to check the validity of the small deformation limit, for which Hertz mechanics is applicable.  The relation is well respected for the stiffest beads and only the yellow gels show a slight departure (Fig.~\ref{fig:scaling}b), \shunsuke{confirming that the distant regions of the beads remain spherical except in the case of the yellow beads.} 

It is now possible to minimize the total energy of the system $U = U_{\mathrm{E}} + U_{\gamma}$, where $U_{\mathrm{E}}$ is the elastic energy and $U_{\gamma}$ is the capillary energy. The capillary contribution is $U_{\gamma}=\gamma S$, where $S$ corresponds to the interfacial area between aqueous phase and the surrounding oil. We {approximate} $\gamma_{\mathrm{drop\!-\!oil}}=\gamma_{\mathrm{gel\!-\!oil}}=\gamma$, since the PEGDA gel is highly water-rich and therefore compositionally close to the liquid phase. 

{For $U_{\mathrm{E}}$ several models have been developed to relate elastic energy to the deformation of contacting spherical particles~\cite{shima_tatara_1993,johnson_adhesion_1997,Lin2009Spherical,Mukhopadhyay2011Packing,liu_capillary_2021}. In practice, however, the classical Hertz contact model remains a reasonable approximation up to $\delta/R\simeq0.2$ for many soft materials, though limitations in accuracy have been reported in some systems~\cite{Dintwa2008Accuracy,Giannopoulos2025Cell}. Since the Hertz model also provides meaningful physical insights, we employ it to describe the elastic response. This yields $U_{\mathrm{E}} = c\,E^{*}\sqrt{R}\,\delta^{5/2}$, where the analytical prefactor for the contact between two spheres is $c=8/(15\sqrt{2})\simeq0.38$.}


Combining the expressions for $U$ with the geometric parameters, the dimensionless total energy becomes
\begin{equation}
\hat{U}(\hat{\delta})=\frac{2c}{E_c}\,\hat{\delta}^{5/2}-4\pi \hat{\delta}+8\pi,
\label{eq:uhat}
\end{equation}
where $\hat{U}\equiv U/(\gamma R^{2})$ is the dimensionless energy and $\hat{\delta}\equiv \delta/R$ is the dimensionless indentation. This energy exhibits a single minimum for every value of $E_c$ (Fig.~\ref{fig:scaling}c). By finding the minimum of $\hat{U}$ and using the corresponding value of $a/R$ from Fig.~\ref{fig:scaling}b, it is possible to \shunsuke{compare this model  with the experimental data, with no free parameters}, as shown by the solid line on Fig.~\ref{fig:scaling}d.

The relation of $a/R$ with $E_c$ can be interpreted in simpler terms by considering the small indentation limit for small values of $E_c$. In this case we can use the approximation $(a/R)^{2}=2\hat{\delta}-\hat{\delta}^{2}\simeq2\hat{\delta}$. Applying the stationarity condition $\partial\hat{U}/\partial\hat{\delta}=0$ then yields the relation
\begin{equation}
\frac{a}{R} \simeq \sqrt{2} \left( \frac{4\pi}{5c} \right)^{1/3} E_c^{1/3},
\label{eq:scaling1}
\end{equation}
\shunsuke{which is represented by the dashed line on Fig.~\ref{fig:scaling}d. The good agreement between the model and experimental data, in the absence of any free fitting parameters, shows that this energy minimization captures the dominant physical balance describing the DEGs. However this simple model cannot account for more complex behaviors that involve nonlinear physical ingredients.}

\begin{figure}[ht]
  \centering
  \includegraphics[width=0.8\columnwidth]{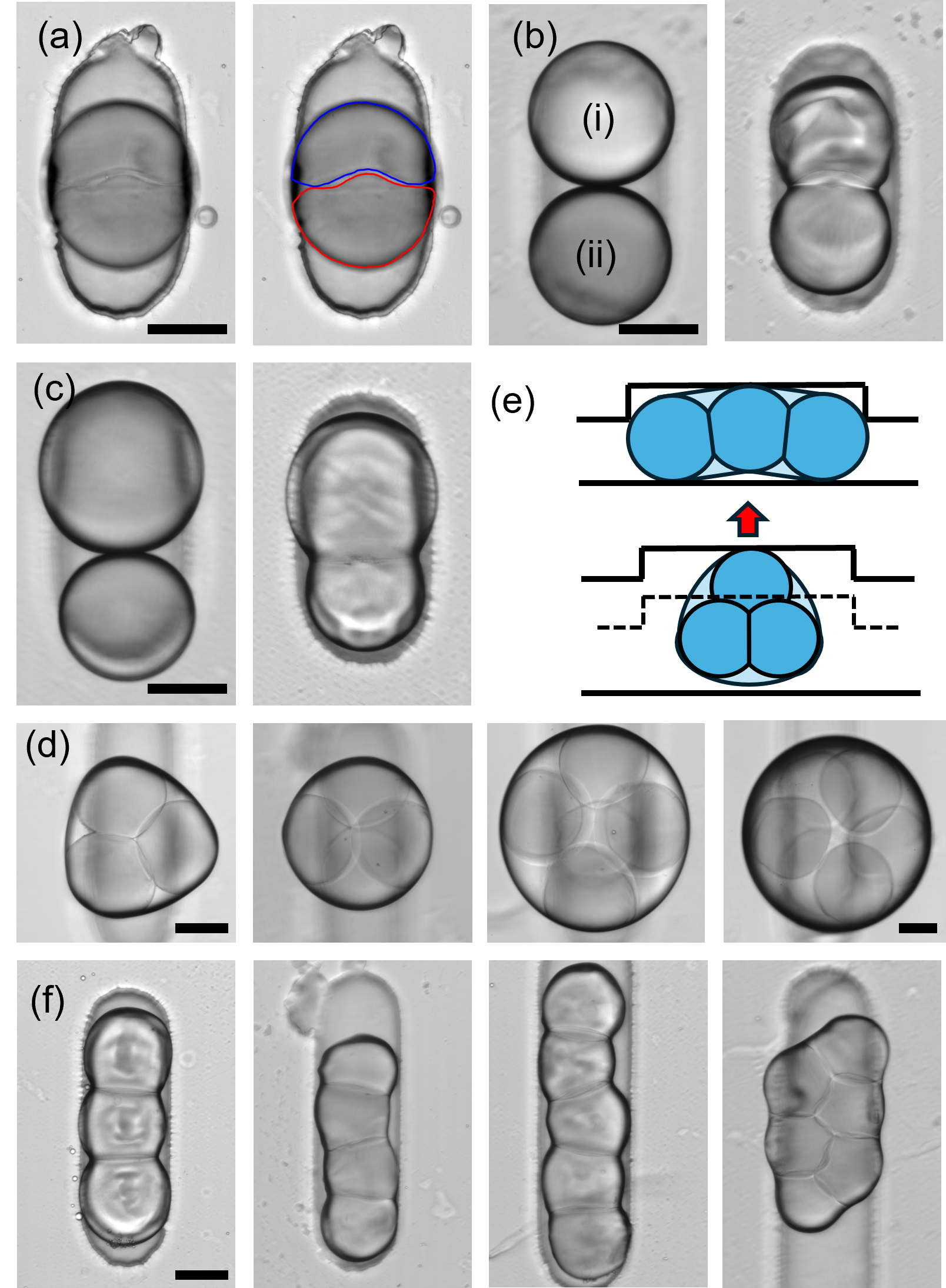}
  \caption{(a) Very soft beads ($E_c > 0.1$) showing buckling in the dry limit ($\phi = 0$). The right image is the same as the left image but with the interface between the two beads highlighted with the colors to help visualization.  
  (b) Pair with unequal stiffnesses:  Yellow bead (i) is soft, and black bead (ii) is stiff, shown before merging (left) and in the dry limit (right). Only the soft bead buckles.
  (c) Pair with unequal sizes shown before merging (left) and in the dry limit (right). 
  (d) $N=3$–6: DEGs in compact arrangements at high $\phi$.  
  (e) Air injection into the cavities raises the chamber ceiling, enabling reconfiguration.  
  (f) $N=3$–6: DEGs at low $\phi$. All scale bars, $100~\mu\mathrm{m}$.}
  \label{fig:multi}
\end{figure}

Indeed the linear theory can break down in several interesting ways. First consider the case of very soft gels, shown in Fig.~\ref{fig:multi}a. These gels, with a PEGDA:PEG ratio of 0.08, had a value of $E^*$ below what could be measured with the microfluidic setup, corresponding to a value of $E_c>0.1$. In this case the surface tension is highly dominant, leading to a nearly spherical outer shape of the bead doublet. As a result the gel beads must undergo very large deformations that lead to the buckling of the contact area between them. Other nonlinear deformations are observed when the two beads have different values of $E^*$ (Fig.~\ref{fig:multi}b). Here the soft bead (i) displays strong wrinkles on its surface, in contrast with the more rigid bead (ii) that remains smooth. Wrinkling is also observed when the beads have different sizes, as shown in Fig.~\ref{fig:multi}c, where the smaller bead displays wrinkles in the distal regions.

Finally, it is also possible to extend the encapsulation of beads beyond the $n=2$ case, to include 3, 4, 5, or 6 beads within a single droplet (Fig.~\ref{fig:multi}d-f). While high liquid fractions always lead to three-dimensional arrangements of the gel beads, the low $\phi$ limit can \shunsuke{yield} either linear arrangements or three-dimensional stacking of the beads, depending on the way the beads are manipulated in the channel. Both the linear and the packed configurations can be preserved when the beads are extracted from the traps, as shown in SI Fig.~S3~\cite{saita_supplementary}.

{In conclusion the current study determines the shape of soft beads subjected to capillary forces by combining microfluidic production, trapping, and mechanical manipulation of aqueous microgels in oil. The resulting relation predicts that the capillary effects induce large deformations in the limits of soft particles or small particles, with implications for the shapes of small microparticles and nanoparticles~\cite{hu_particle_2020} and material printing~\cite{Pawar2012Arrested,Roh2017Capillary}. Moreover the co-encapsulation of multiple gel beads (Fig.~\ref{fig:multi}) shows linear or compact arrangements that depend on the bead friction and deformations. These arrangements, which cannot be obtained by co-encapsulating liquid droplets~\cite{guzowski_droplet_2015,li2017new}, could be leveraged for microtissue engineering~\cite{rojek_microfluidic_2022}.}

{Beyond elastocapillarity, the energy minimization in the dry limit recalls the equilibrium between elasticity and adhesion. This interaction has been quantified using a range of theories, such as the Johnson-Kendall-Roberts (JKR) theory that equilibrates elastic deformation and adhesion energy at the contact interface~\cite{johnson_adhesion_1997}. 
As such the arguments developed in Eqs.~\eqref{eq:uhat} and \eqref{eq:scaling1} can readily be extended to yield the relationship of $a/R\sim\hat{E}_c^{1/3}$, where $\hat{E_c}$ is the elastocapillary number based on surface adhesion, with interesting applications namely to the shape of adhesive cells and multicellular agregates~\cite{chu_jkr_2005,Gonzalez-Rodriguez2012}.}

\textit{Acknowledgements--} The authors acknowledge microfabrication support by Caroline Frot (Ladhyx) and by the X-Fab of Ecole Polytechnique. Helpful discussions are also acknowledged with Leon Gebhard and Hiba Belkadi. This work was partially funded by the European Union (ERC-2023-ADG grant number 101142018 MELCART) and from the French National Research Agency (ANR) as part of the France 2030 plan under grant agreement ANR-22-EXES-0013.

\textit{Data availability—} The data that support the findings of this article are openly available on Zenodo at \url{https://zenodo.org/records/22706445}.

\bibliographystyle{apsrev4-2}
\bibliography{DEGs_PRL}

\end{document}


\title{Deformation and organization of droplet-encapsulated soft beads}

\author{Shunsuke Saita}
\affiliation{\ladhyx}
\affiliation{\kyotomech}

\author{Finn Bastian Molzahn}
\affiliation{\ladhyx}

\author{Clara Delahousse}
\affiliation{\ladhyx}
\affiliation{\pasteur}

\author{Julien Husson}
\affiliation{\ladhyx}

\author{Charles N. Baroud}
\email{charles.baroud@polytechnique.edu}
\affiliation{\ladhyx}
\affiliation{\pasteur}

\date{\today}

\begin{abstract}
\end{abstract}

\maketitle
\clearpage
\onecolumngrid
\appendix

\renewcommand{\thefigure}{S\arabic{figure}}

\section{Homogeneity of the cross-linking in the beads}

We estimate the oxygen inhibition using the scaling relations of Dendukuri et al.\cite{dendukuri_modeling_2008}.

The Damköhler number is defined as
\begin{equation}
    Da = Q_E\varepsilon[PI]I_0H^2 / (D_0[O_2,eq]),
\end{equation}
where $Q_E$ is the quantum yield of formation of initiating radicals, $\varepsilon$ is the molar absorptivity of the photoinitiator, $[PI]$ is the photoinitiator (Darocur1173) concentration, $I_0$ is the UV light intensity, $H$ is the characteristic length scale of the system, $D_0$ is the oxygen diffusivity, and $[O_2,eq]$ is the equilibrium oxygen concentration. Since the material used here is the same as Dendukuri et al, the parameters $Q_E$, $\varepsilon$, and $[O_2,eq]$ are assumed to be the same as in that study. 

In our system, the characteristic length scale corresponds to the droplet diameter used for polymerization, D = 174 µm(Large), 106(Small) in median. In their experiments, a 100 W HBO mercury lamp with a 365 nm i-line filter was used, while in our setup, the gel beads were polymerized using a Nikon C-LHGFI HG mercury lamp (130 W) with a DAPI fluorescence filter at ND = 0 (maximum output). Although the absolute UV intensity $I_0$ was not directly measured, it is therefore expected to be comparable to, or larger than, the value reported in that study. The Damköhler number $Da\simeq1700$ for large beads, $Da\simeq600$ for small beads. Using the relations $\tau_i \sim 1/Da$ and $\delta_{i,c} \sim  Da^{-1/2}$, and replacing the geometric and chemical parameters with our experimental values, the dimensionless induction time can be converted to dimensional time using $t=\tau H^2/D_0$. This gives $ti \simeq 0.6$~s (large) and 1.8~s (small), both much shorter than our UV exposure time ($t=5$~s). The oxygen-inhibited layer thickness is estimated to be $\delta_{i,c}\times H\simeq4.2~\mu$m for both the large and small beads, which is much smaller than the bead diameter. Crosslinking is therefore expected to occur throughout most of the bead volume, resulting in an approximately spatially uniform modulus.



\clearpage

\section{Supplementary figures}

\begin{figure}[h]
    \centering
    \includegraphics[width=0.5\linewidth]{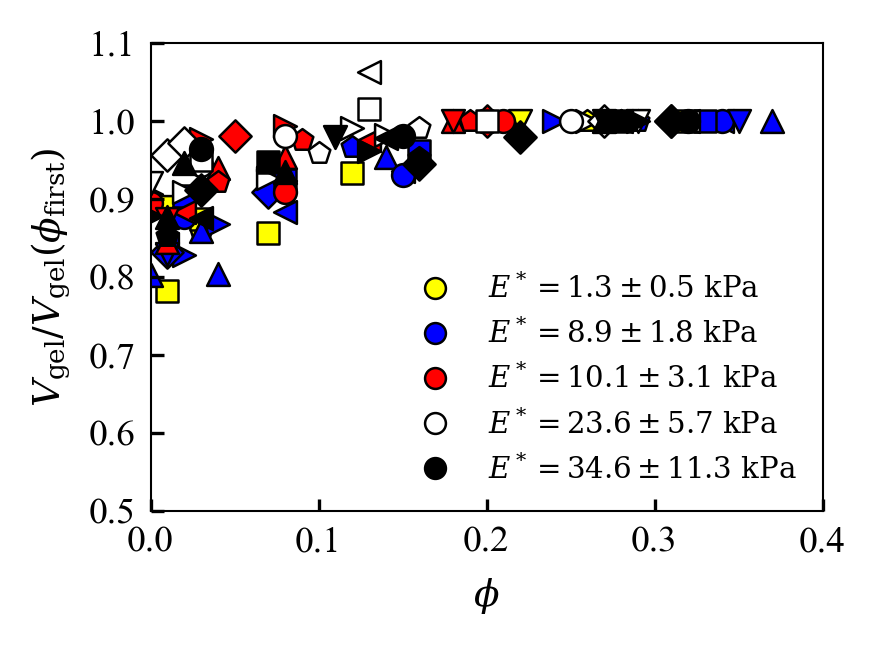}
    \caption{Gel volume evolution with decreasing liquid volume fraction $\phi$.
    Only DEGs that merged without an initial volume loss at the onset of coalescence were included. For these samples, the liquid volume fraction at this initial state was defined as $\phi_{first}$, and $\phi$ was subsequently reduced from this reference: black ($n=8$), blue ($n=8$), red ($n=7$), white ($n=7$), and yellow ($n=3$).}
    \label{fig:placeholder}
\end{figure}


\begin{figure}[h]
  \centering
  \includegraphics[width=0.5\columnwidth]{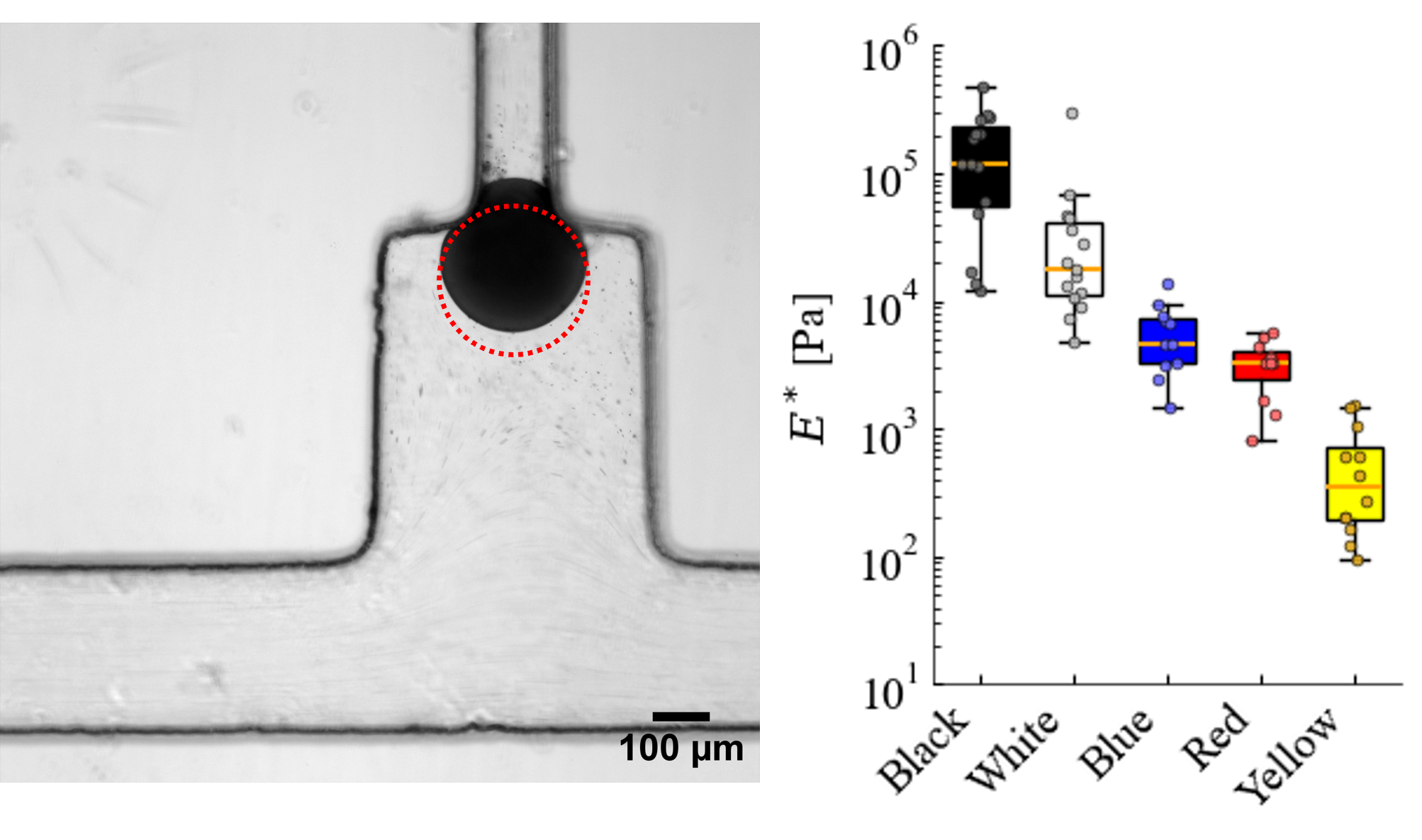}
  \caption{Measurement of effective elasticity $E^{*}$ of using a microfluidic deformation assay. Left: Image of a bead in the microfluidic constriction. Right: Values obtained using the formalism of Ref.~\cite{moore_clogging_2023}.  
}
  \label{fig:SI_props}
\end{figure}
    



\begin{figure}[h]
  \centering
  \includegraphics[width=0.5\columnwidth]{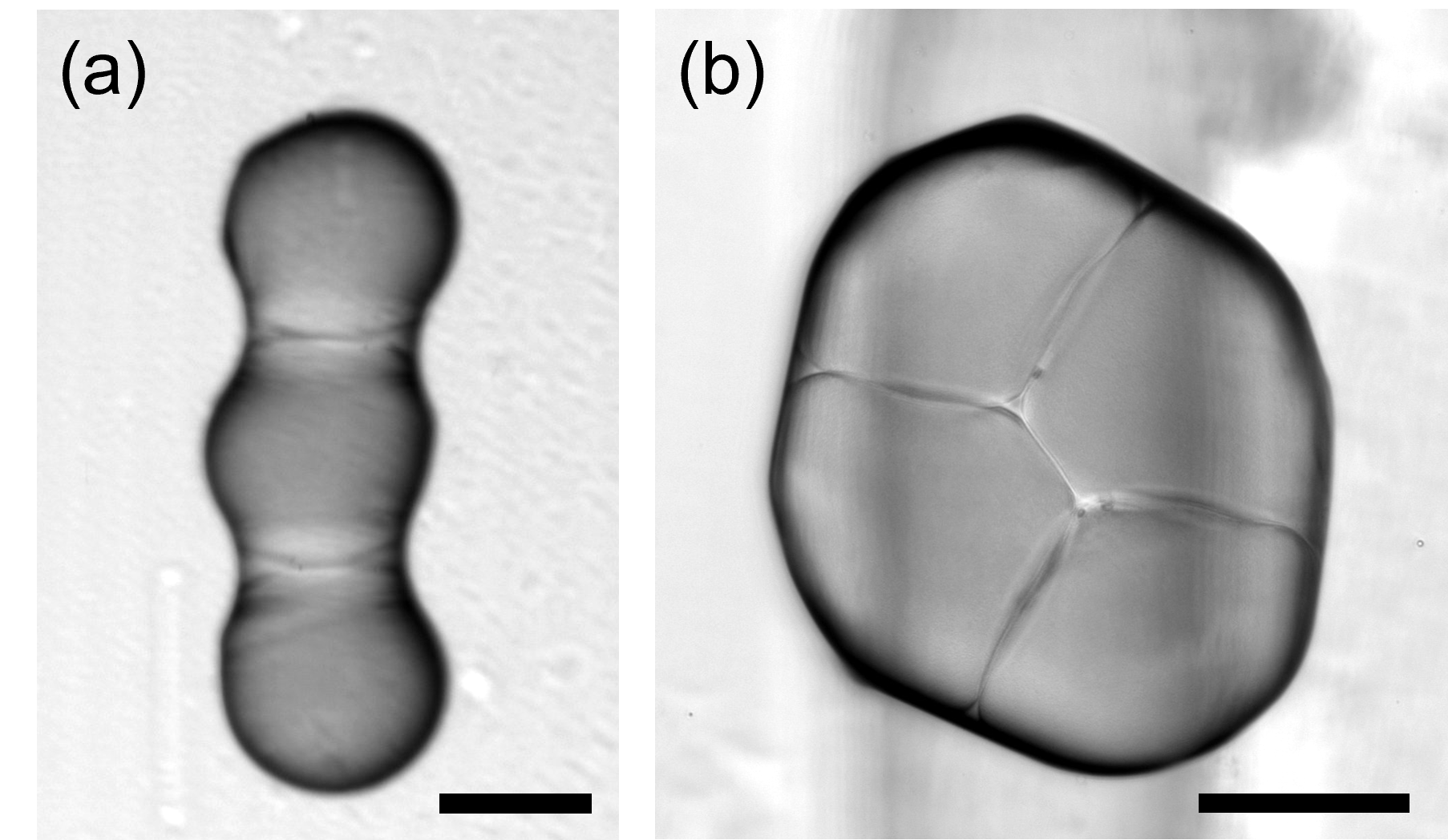}
  \caption{
\textbf{Linear or aggregated configurations can be stable outside the microfluidic traps} (a) $N=3$: DEGs in linear arrangements at low $\phi$ is stable out of trap. (b) $N=4$: DEGs in compact arrangement at low $\phi$. All scale bars  $100~\mu\mathrm{m}$.}
  \label{fig:multi}
\end{figure}





\clearpage
\newpage

\bibliographystyle{apsrev4-2}
\bibliography{DEGs_PRL}